\documentstyle[prl,aps,epsf]{revtex}
\begin{document}
\draft

%\twocolumn[\hsize\textwidth\columnwidth\hsize\csname@twocolumnfalse%
%\endcsname

\title{Coherent Control of Photocurrents in Graphene and Carbon Nanotubes}
\author{E.J. Mele$^1$, Petr  Kr\' al$^2$ and David Tom\'anek$^3$}
\address{$^1$Department of Physics and Laboratory for Research on the Structure of
 Matter\\ 
University of Pennsylvania, Philadelphia PA 19104 \\
$^2$Department of Physics, University of Toronto, Toronto Ontario M5S 1A7 Canada \\ 
 $^3$Department of Physics and Astronomy, Michigan State University, East Lansing MI 48824 }
\date{\today}
\maketitle

\begin{abstract}
Coherent one photon ($2 \omega$) and two photon ($ \omega$) electronic excitations
 are studied for  graphene sheets and for carbon nanotubes
 using a long wavelength theory for the low energy electronic
 states. For graphene sheets we find that coherent superposition of these
 excitations produces a polar asymmetry in the momentum space distribution of the
excited carriers with an angular dependence which depends on the relative polarization and phases of the incident fields. For semiconducting nanotubes
 we find a similar effect which depends on the square of the
 semiconducting gap, and we calculate its frequency dependence. 
 We find that the third order nonlinearity which controls the direction of the photocurrent is robust for semiconducting tubes and   vanishes in the continuum theory for conducting
 tubes. We calculate corrections to these 
 results arising from higher order crystal field effects on the
 band structure  and briefly discuss some applications of the theory. 
\end{abstract}
\pacs{PACS: 42.50.Hz,42.65.Sf,61.48+c,78.20.Jq} 
%\vskip -0.5 truein
%]

\section{Introduction}

The magnitude and direction of photocurrents in semiconductors are ordinarily controlled
using applied bias voltages. Interestingly the direction of a photocurrent in a
 semiconductor can also be controlled
 without bias voltages through phase coherent control of the incident optical fields.  In a typical experiment
 an initial and final state are simultaneously coupled using two coherent
excitations:  one  photon excitation at frequency
 $ 2 \omega$ and two  photon excitation at $ \omega$. The coherent superposition of these
 two excitations can lead to a polar asymmetry in the momentum space distribution of the excited photocarriers
 and therefore to a net photocurrent. The effect has been discussed theoretically \cite{shap,atan} and observed
 experimentally in photoyield from atoms \cite{yin} and for photocurrents in 
semiconductors \cite{dupont}. Recently, two of us have proposed that for carbon 
 nanotubes  this effect could provide directional control of
 a photocurrent along the tube axis \cite{kral1} and even suggests a  novel method for biasing  the diffusion of
 ionic species which intercalate within the  nanotubes. 

 In this paper we study the excitations which lead to this 
 effect both for graphene sheets and for carbon nanotubes. In both these systems the low energy 
 electronic properties relevant to most solid state effects are determined by an interesting feature of the band structure. The isolated graphene
 sheet has only an incipient Fermi surface; it is actually a zero gap semiconductors where the
 conduction and valence bands meet at precise points in momentum space. The carbon nanotube
 is a cylindrical tubule formed by wrapping a graphene sheet and  for metallic tubes  the ``zero gap" feature
 manifests itself in a peculiar doubling of the low energy electronic  spectrum with ``pairs" of
 forward and backward moving excitations at both $k_F$ and $-k_F$ \cite{km}. In either case
the low energy electronic spectra are described by a two component Dirac Hamiltonian \cite{km}.

 In this paper we
 develop the theory of phase coherent one- and two- photon excitation within this model. 
 The application to the graphene sheet turns out to be a useful pedagogical model which 
 is unusual for a semiconductor  and nicely illustrates the
 origin of phase coherent control of photocurrents for a graphene derived system. 
  For graphene it is inappropriate to analyze the third order nonlinearity by 
 analogy with the third order response in 
 atomic systems, as has been done previously for semiconductors\cite{dupont}.   
  Instead we find that the third order response probes  the rather unique geometry of the extended low energy electronic eigenstates which occur within the  graphene sheet.  The application of
 the model to a carbon nanotube shows, interestingly,  that the third order nonlinearity is suppressed for excitations
 between the lowest subbands of any conducting nanotube and 
vanishes completely for transitions between the lowest subbands of a  conducting ``armchair" tube, but it is nonzero and robust for the gapped subbands
 of a semiconducting tube. In fact the effects we calculate are significantly stronger for
 semiconducting nanotubes than for a conventional semiconductor.  
 In principle this effect might be used to distinguish conducting and
 semiconducting tubes in a compositionally mixed sample. Other possible applications of the
 idea will be discussed later in the paper.  

In this paper we briefly review the effective mass theory for the graphene
 sheet in Section II. In Sections III and IV we derive the interaction
 terms in the long wavelength theory which couple the electrons to time
varying electromagnetic fields and  present a calculation of the
 coherent  third order nonlinear optical excitations using this model.
 Section V applies the results to study the third order response of
 an isolated infinite graphene sheet. In Section VI we use the theory
 to study third order effects for conducting and semiconducting carbon
 nanotubes.  A discussion and some applications of the results are
 presented in Section VII. 

\section{Effective Mass Theory}

  In this section we briefly review  the effective mass description of the low energy electronic states. The theory is developed
 for an ideal graphene sheet, a section of which is shown in Figure 1. The primitive cell of this structure contains two atoms, labelled A and B in the figure. The lattice is unchanged after
  a translation by any combination of the two primitive translations vectors
\begin{eqnarray}
\vec{T}_1 &=& a(1,0) \nonumber\\
\vec{T}_2 &=& a(\frac{1}{2},\frac{\sqrt{3}}{2})
\end{eqnarray}
where the bond length $d = a/\sqrt{3}$. We introduce a pair of primitive translation vectors
for the reciprocal lattice $\vec{G}_i$ such that $\vec{G}_i \cdot \vec{T}_j = 2
\pi \delta_{ij}$,
 yielding
\begin{eqnarray}
\vec{G}_1 &=& \frac{4 \pi}{\sqrt{3}a} (\frac{\sqrt{3}}{2},-\frac{1}{2}) \nonumber\\
\vec{G}_2 &=& \frac{4 \pi}{\sqrt{3}a} (0,1)
\end{eqnarray}
which generate a triangular lattice in reciprocal space.
\begin{figure}
   \epsfxsize=3.0in
   \centerline{\epsffile{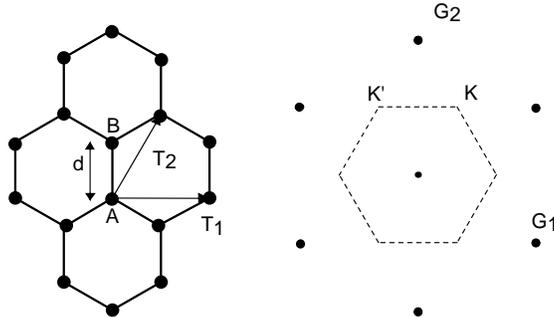}}
   \caption{Direct and reciprocal space structures of the graphene lattice. The primitive
 cell contains two sublattice sites labelled A and B in the left panel. The right panel shows
 the first star of reciprocal lattice vectors and the first Brillouin zone. The long wavelength
 theory expands the electronic Hamiltonian for momenta near the K and K' points at the Brillouin
 zone corners.}
\end{figure}
 The critical points K(K') are important to 
  our discussion, and they occur at the corners of the Brillouin zone of this reciprocal
 lattice at the positions
\begin{eqnarray}
K &=& \frac{1}{3} (\vec{G}_1 +  2 \vec{G}_2) \nonumber\\
  &=& \frac{4 \pi}{ 3 a} (\frac{1}{2}, \frac{\sqrt{3}}{2}) \nonumber\\
 \nonumber\\
K' &=&  \frac{1}{3} (- \vec{G}_1 +  \vec{G}_2) \nonumber\\ 
  &=& \frac{4 \pi}{ 3 a} (-\frac{1}{2}, \frac {\sqrt{3}}{2})
\end{eqnarray}
The ``bonding" and ``antibonding" $\pi$ electron bands meet precisely at these K(K') points in
 reciprocal space. This band touching is required by symmetry for this system and it is correctly described by the
 simplest model for electron propagation within the graphene sheet which is a tight binding model in
 which the hopping of an electron between neighboring sites is set by a single energy, t. 
Thus we have
\begin{eqnarray}
h_{\mu \nu} = \langle \phi_{\mu} | H | \phi_{\nu} \rangle  = &t& \,\,(\rm{nearest neighbor}\, \mu \nu) \nonumber\\
							   &0&  \,\,(\rm{otherwise})
\end{eqnarray} 
Working in the sublattice basis and at crystal momentum $\vec{k}$ we have the Hamiltonian
\begin{equation}
H(\vec{k}) = t  \left( \begin{array}{cc} 0&1 +e^{-i \vec k \cdot \vec T_2} + e^{-i \vec k \cdot (\vec T_2 
    - \vec T_1)} \\
  1 +e^{i \vec k \cdot \vec T_2} + e^{i \vec k \cdot (\vec T_2 - \vec T_1)}&0 \end{array} \right)
\end{equation}
If we set $\vec k = \vec K + \vec q$ and expand the Hamiltonian for $qa << 1$ we obtain
the long wavelength Hamiltonian
\begin{eqnarray}
H_K (\vec q) &=& \frac{\sqrt{3}ta}{2} \left( \begin{array}{cc} 0&q_x + i q_y \\
							q_x - i q_y&0 \end{array} \right) \nonumber\\
	     &=&  \hbar v_F \vec {\sigma}^* \cdot \vec {q}
\end{eqnarray}
where $\vec {\sigma}$ are the $2 \times 2$ Pauli matrices.  A similar expansion near the K' point yields
\begin{equation}
H_{K'} =  - \hbar v_F \vec {\sigma} \cdot \vec {q}
\end{equation}
 Identifying each of
 the critical points with the index $\alpha$ so that $\alpha=1$ denotes
 the K point and $\alpha = -1$ denotes the K' point, these
 Hamiltonians can be rotated into diagonal form with the unitary operators
 \begin{equation}
 U_{\alpha}  (\vec q) =  
 \frac{1}{\sqrt{2}} \left( \begin{array} {cc} 1&1 \\
				       -\alpha e^{-i \alpha \theta} &
					\alpha e^{-i \alpha \theta}
\end{array} \right)
\end{equation} 
where $\theta = \tan^{-1} (q_y / q_x)$. Thus
\begin{equation}
U_{\alpha}^{\dagger}(\vec q) H_{\alpha} (\vec q) U_{\alpha}(\vec q) =
 \hbar v_F \left( \begin{array} {cc} -|q|&0 \\ 0&|q| \end{array} \right)
\end{equation}
so that equations (6) and (7) describe pairs of bands which disperse linearly
 away from the critical K and K' points. Note also that $H_{\alpha} (\vec q) =
 H_{\alpha}^* (- \vec q)$ as expected. Equations (6) and (7) and the unitary
 rotations in equation (8) provide the appropriate description of all the low energy
 electronic excitations required for this problem. 

\section{Graphene-Field Interaction Hamiltonian}

In this section we collect several results we need to  describe the coupling of electrons described
 by Equations (6) and (7) to the electromagnetic potentials. For a particle of charge $Q$
 interacting with the electromagnetic vector potential $\vec A$ and scalar potential
 $\Phi$ the momentum and energy are shifted $\vec p \rightarrow \vec p - Q \vec A /c$
 and $E \rightarrow E - Q \Phi$\cite{scull}. Thus the interaction Hamiltonian which couples the
 Dirac particle to the time varying vector potential $(A_x,A_y)$ is
 \begin{equation}
 H_{\alpha,int} =   - \alpha  v_F Q (A_x \sigma_x -\alpha A_y \sigma_y)/ c
 \end{equation}
 Note that this interaction  operator can also be obtained by calculating the velocity operator as  the commutator of
 the position operator with the unperturbed Hamiltonian:
 \begin{equation}
 \vec v_{\alpha} = \frac{i}{\hbar} \left[ H_{\alpha}, \vec r \right]
 \end{equation}
 and therefore
 \begin{equation}
 \vec v_{\alpha} =  \alpha v_F (\sigma _x,- \alpha \sigma _y)
 \end{equation} 
 so that  $H_{\alpha,int} = - \vec j_{\alpha} \cdot \vec A /c$. 

 It is useful to rotate the interaction Hamiltonian (10) into the band basis using the
 unitary  operators in  equation (8). To do this we write  $\hat q =  (q_x ,  q_y)/|\vec q| $  and compute $H^b _{\alpha,int} = U^{\dagger}_{\alpha}(\hat q) H_{\alpha,int}(\hat q) U_{\alpha}(\hat q)$ giving
\begin{equation}
H^b _{\alpha,int}(\hat q,\vec A)  = \frac { e v_F}{c} \left(  \begin{array} {cc} -{ \hat q \cdot \vec A}&{-i \alpha \hat z \cdot (\hat q \times \vec A)} \\ { i \alpha \hat z \cdot ( \hat q \times \vec A)}&{ \hat q \cdot \vec A} \end{array} \right) 
\end{equation}
This demonstrates that the coupling between the Bloch  electrons and the vector potential depends
 on the angle between $\vec v$ and $\vec A$ and that the interband matrix elements
 (which are the off diagonal terms in Equation (13)) vanish when the two are collinear. Indeed, $H_{\alpha} (\vec q)$ and $H_{\alpha,int}(\hat q,\vec A)$ commute
 along these special lines in reciprocal space, so that interband transitions
 are forbidden along this trajectory.  This  peculiar feature can
 be traced to the absence of a mass term in the effective Hamiltonians in equations (6)
 and (7) which would ordinarily mix the plane wave solutions to (6) and (7) and thereby
 allow interband transitions by coupling with the long wavelength current operator.  When
 $\vec v$ and $\vec A$ are not collinear interband transitions are allowed and the
 transition amplitudes are fixed  by the mismatch in their orientations in the graphene plane.
 This has interesting consequences for possible coherent control of nonlinear optical processes
 in the nanotubes, as we show below. 

 \section{Nonlinear Optical Excitations}

 In this section we present a calculation of the transition probabilities for  the third order nonlinear optical excitations
 among the electronic states given by the models in Sections II and III. We introduce
 time varying fields of the form
 \begin{equation}
 \vec A(\vec r, t) = \vec A_{\omega} e^{- i \omega t + i \phi_1} + \vec A_{2 \omega} e^{- 2i \omega t + i \phi_2} + c.c.
 \end{equation}
 and study the response of the system to third order in these exciting fields.  Asymmetries
 in the photocurrent are controlled by the coherent excitation of electrons from an initial state to
 a final state by one photon ($2 \omega$) and by two one  photon ($ \omega$) processes. 
 The coherent mixing of these two processes is studied  by evolving the density matrix
 to third order in the exciting fields and isolating the terms proportional to $A_{\omega}
 A_{\omega} A_{-2 \omega}$.

 It is convenient to study the time evolution of the one particle density matrix $ \rho = \langle \Psi^{\dagger}(\vec r) \Psi(\vec r') \rangle$.  The Hamiltonian for our system is $H_{\alpha} + H_{\alpha,int}$ and we
 work in the Heisenberg representation so that
 \begin{equation}
 \frac{d \rho}{dt} = \frac {i}{\hbar} \left[ H_{\alpha,int} (t) , \rho \right]
 \end{equation}
 where $H_{\alpha}$ is the free particle Hamiltonian and $H_{\alpha,int}(t) = e^{i H_{\alpha} t} H_{\alpha,int} e^{-i H_{\alpha} t}$. In the band basis the density matrix in the initial state has the form
 \begin{equation}
\rho_0 =  \rho(t = - \infty) = \left( \begin{array} {cc}1&0 \\ 0&0 \end{array} \right)
 \end{equation}
 since only the negative energy states of the Hamiltonians (6) and (7) are initially occupied. 

 	We expand the density matrix order by order in the exciting fields
\begin{equation}
\rho(t) = \rho_0 + \rho_1 + \rho_2 + \rho_3 + ... 
\end{equation}
Integrating equation (17) to first order in the applied fields gives
\begin{equation}
\rho_1 = \frac{iev_F}{ \hbar c} \left( \begin{array} {cc} 0 & F_1(t) \\ - F^*_1(t)& 0
			        \end{array} \right)
\end{equation}
where
\begin{eqnarray}
F_1(t)&=& \frac{\alpha \hat z \cdot (\hat q \times \vec A_{\omega}) e^{-i (\Delta + \omega) t +i \phi_1}}
             {-\Delta -\omega -i \delta} + 
	\frac{ \alpha \hat z \cdot (\hat q \times \vec A_{- \omega}) e^{- i (\Delta -  \omega) t - i \phi_1}}
             {-\Delta +\omega -i \delta} \nonumber\\
&+&  \frac{\alpha \hat z \cdot (\hat q \times \vec A_ { 2\omega}) e^{-i (\Delta +  2 \omega) t +i \phi_2}}
             {-\Delta - 2 \omega -i \delta} +
	 \frac{\alpha \hat z \cdot (\hat q \times \vec A_{- 2\omega}) e^{- i (\Delta -  2 \omega) t -i \phi_2}}
             {-\Delta + 2 \omega -i \delta}
\end{eqnarray}  
where $\Delta = 2v_F q$ and $\delta$ is a positive infinitesimal. 

The second order terms $\rho_2$ include the lowest order changes to the occupation probabilities
 which can be induced with excitation by the $\omega$ or the  $2 \omega$ fields
\begin{equation}
(\dot {\rho}_2)_{22} =  - (\dot {\rho}_2)_{11} = \frac { 2 \pi e^2 v^2 _F} {\hbar^2 c^2}( 
  | \alpha \hat z \cdot (\hat q \times \vec A_{\omega})|^2 \delta(\Delta - \omega) +  
  |\alpha \hat z \cdot (\hat q \times \vec A_{ 2 \omega})|^2 \delta(\Delta -    2\omega))
\end{equation}
as well as  oscillating nonlinear off diagonal coherence terms
\begin{equation}
(\rho_2)_{12} = - \frac {i e^2 v^2_F}{\hbar^2 c^2} F_2(t)
\end{equation}
Anticipating the situation $2 \omega \approx \Delta$, the most important contribution to $F_2(t)$ has the form
\begin{equation}
F_2(t) =  \frac{ (\hat q \cdot \vec A_{- \omega})( \alpha \hat z \cdot \hat q \times \vec A_{- \omega}) e^{-i(\Delta - 2 \omega) t -2i \phi_1}}
	       {(- \Delta + \omega - i \delta)(- \Delta + 2 \omega - i \delta)}
\end{equation}
The second order coherence term in equation (22) leads to a transition rate which is third order in
the applied fields and is the source of the polar asymmetry of the photocurrent. 
\begin{equation}
(\dot \rho_3)_{22} = \frac {8 \pi e^3 v^3_F}{ \Delta \hbar ^3 c^3}
    \, {\rm Re} \,  (( \alpha \hat z \cdot \hat q \times \vec A_{ 2 \omega}) (\hat  q \cdot \vec A_{- \omega})( \alpha \hat z \cdot \hat q \times \vec A_{- \omega}) e^{i (\phi_2 - 2 \phi_1)})
 \delta(\Delta - 2 \omega)
\end{equation}
Equation (23)  contains a factor of two from the sum over the (physical) spins.
 Equation (23) presents the main result of the paper. It shows that the transition rate depends on the polarization and phases of both exciting fields and the Bloch wavevector 
$\vec q$. We will explore the consequences of the geometric structure of this result for the graphene
 sheet and for carbon nanotubes in the following two sections. For the moment we note that
 the result is odd in the  direction of the Bloch wavevector $\hat q$ and even in the critical point index $\alpha$ (it depends on $\alpha ^2$) and  therefore
 the symmetry breaking nonlinearity is nonzero after integration over the full Brillouin zone. 

\section{Application to Graphene}

In this section we apply the formalism developed in section III to study the coherent
 optical control of photocurrents for a single graphene sheet. The model nicely illustrates the
 selection rules which apply  in this geometry, and the results can be extended to analyze the
 more complex situation for the nanotube which will be presented in section VI.

 We note that both equations (20) and (23) contain terms which describe transitions from the valence
 to the conduction band at the frequency $2 \omega = \Delta$. Equation (20) is the
 ordinary linear absorption in the material. Interestingly we see that the angular
 distribution of the excited photocarriers is not isotropic but rather follows a $\rm sin^2 \phi$
 dependence with respect to the polarization of the exciting radiation. Nevertheless this
 angular distribution has even parity and thus does not produce a net current. On the other hand equation (23)
 gives an angular distribution that breaks the inversion symmetry of the graphene sheet.
 The symmetry breaking is actually implicit in the coherent superposition of the exciting
 fields. We will estimate the prefactors to compare the relative strengths of these terms for accessible
 laboratory fields later in the paper; for the moment we note that the nonlinear terms in 
 equation (23)  typically contribute $\approx 10^{-3}$ of the total transition rate, and
 thus the induced anisotropy while nonzero (and we believe  measurable) is a
 subtle effect. 

 The angular distribution in equation (23) is controlled by the  relative polarizations and
 phases of the incident $\vec A_{\omega}$ and $\vec A_{2 \omega}$ fields. Figure 2 displays polar
 plots of the angular
 distributions for the situation where the amplitudes are in phase (i.e. $\vec A_{\omega}$ and $\vec A_{2 \omega}$ in equation (14) are presumed to be real) for various incident polarizations.
 Note that the underlying electronic dispersion relations are completely isotropic
 in the linearized theory, and thus only the relative polarization of the
 two exciting fields is relevant for the interference pattern.  In all cases  $\vec A_{\omega}$ is taken to be polarized along the horizontal direction
 shown by the arrow in the plots. In each
 plot we observe a node in the current distribution along this direction. This follows from the
 symmetry of $H^b_{\alpha, int}$ in equation (13) which shows that interband coupling is prohibited
 for $\hat q$ parallel to $\vec A$. Nonetheless, the situation for collinear $\omega$ and $2 \omega$
 excitation clearly shows the asymmetry between the ``forward" and ``backward" distribution of the
 photocurrent. The situation is more interesting when the exciting fields are noncollinear. We observe
 that the angular distribution develops a ``three" lobe structure. Ultimately when the
 exciting fields area mutually orthogonal, we recover the ``two lobe" pattern with the angular
 distribution rotated by $\pi / 2$ with respect to the polarization of the incident $\omega$ field. 
 It is useful to quantify the anisotropy of the distribution  by calculating 
 the average polarization of the net photocurrent $\langle \cos \phi \rangle$
 and $\langle \sin \phi \rangle$ averaged over this distribution. One finds
\begin{eqnarray}
\langle \cos \phi \rangle &=& \frac{1}{2} \cos \theta \nonumber\\
\langle \sin \phi \rangle &=& - \frac{1}{2} \sin \theta
\end{eqnarray}
 so that when the $2 \omega$ field is tipped by an angle $\theta$ with respect to the
 $\omega$ field, the photocurrent is oriented in the direction $-\theta$. 
 Finally,  the ``sign" of the effect is determined by the relative phases of the
two exciting fields.  Note that the  phase delays $\phi_1$ and
 $\phi_2$ in the exciting fields of equation (14) modulate the transitions rates\cite{dupont} in
 equation (23) in the combination
\begin{equation}
(\dot \rho'_3)_{22} \rightarrow (\dot \rho_3)_{22} \cos (\phi_2 - 2 \phi_1)
 \end{equation}
 This does not change the qualitative features of the angular distribution but it
can modify both its magnitude and its sign.  Thus the angular distribution in the collinear case
 $\Delta \theta = 0$ can be reversed by advancing the phase of the $\omega$ fields by $\pi / 2$. 
\begin{figure}
   \epsfxsize=5.0in
   \centerline{\epsffile{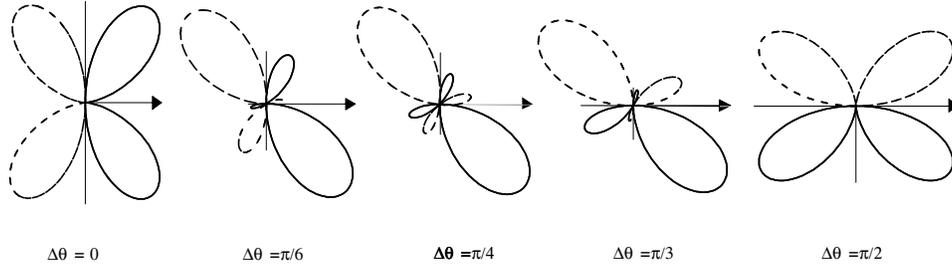}}
   \caption{Angular distributions of the transition rates given by equation (23). In each
panel the $\vec A_{\omega}$ is polarized along the horizontal direction (the direction of
 the arrow in each plot) and $\Delta \theta$ is the angle between the $\vec A_{2 \omega}$ and
  $\vec A_{\omega}$ fields. The polar plot gives the transition rate as a function of the
 angle of the Bloch wavevector $\hat q$ with respect to the direction of $\vec A_{\omega}$.
 These angular distributions are superimposed on the angular distribution
 for  the direct transition rate which is given by the second term in equation (20). 
 The solid curves correspond to $(\dot \rho_3)_{22} > 0$, dashed curves to
 $(\dot \rho_3)_{22} < 0$ when $ \cos (\phi_2 - 2 \phi_1) > 0$.} 
\end{figure}

\section{Application to Nanotubes}

\subsection{Low Energy Theory for Semiconducting Tubes}

In this section we apply the formalism of section III to study phase coherent
 control of a photocurrent on a carbon nanotube. The essential complication
 is that the wrapped structure of the nanotube quantizes the allowed crystal
 momenta so that the transition rate automatically contain an intrinsic
 anisotropy. Nevertheless, the formalism developed in  section III can be extended
 to include this situation.

 We first define the geometry for the single wall nanotube. The nanotube
 is a cylinder formed by wrapping graphene sheet and  the wrapping can be defined by the graphene superlattice translation vector around the tube waist.
  We adopt
 the primitive vectors of equation (1) as a basis and represent the superlattice translation
 $\vec T_{MN}$ as
 \begin{equation}
 \vec T_{MN} = M \vec T_1 + N \vec T_2 = (M + \frac{N}{2},\frac{\sqrt{3} N}{2}) a
\end{equation}
It is useful to define two unit vectors defining the longitudinal and azimuthal
 directions within the graphene plane
\begin{eqnarray}
\hat e_l &=& (\cos \theta _c , \sin \theta _c ) \nonumber\\
\hat e_a &=& (-\sin \theta _c , \cos \theta _c )
\end{eqnarray}
where $\theta _c = \cos ^{-1} (M+N/2)/(\sqrt{M^2 +N^2 +MN})$ is the chiral angle of the tube. 
The wrapping of the tube quantizes the allowed momenta along the
 azimuthal direction $\vec k \cdot \hat e_a = 2 \pi n/(a \sqrt{M^2 + N^2 +MN})$ while the electrons obey free particle boundary conditions along
 the tube direction and the longitudinal component  $\vec k \cdot \hat e _l$ can take any value \cite{mint,ham,sai}. Thus the loci of allowed momenta are ``lines" in reciprocal
 space. These  lines need not intersect the critical K and K' points which
 are used as a reference for the long wavelength theory. To determine the
 mismatch between the allowed crystal momenta and the K and K' point wavefunctions we resolve the Bloch wavevector at K and K' into its longitudinal and azimuthal
 components. We find
 \begin{equation}
 K_{\alpha} = \frac {4 \pi}{3a} (\frac {\alpha}{2}, \frac{\sqrt{3}}{2})
 \end{equation}
 and
 \begin{equation}
 K_{\alpha} \cdot \hat e_a = \frac {2 \pi}{ 3a} \left(
 \frac {2 \alpha M + N ( 3 + \alpha)}{2 \sqrt{M^2 +N^2 + MN}} \right)
 \end{equation}
 which lies along the locus of allowed wavectors when 
 $2 \alpha M + (3 + \alpha) N =  6n$. One third of the $(M,N)$ tubes satisfy this condition, 
 and for the remaining two-thirds of the tubes the K(K') momenta are mismatched
to the kinematically allowed momenta by a mimimum amount
 \begin{equation}
 \Delta_{\alpha} = \frac {2 \pi}{ 3a \sqrt{M^2 + N^2 + MN} } 
                   (-1)^{{\rm mod}(2 \alpha M + (3 + \alpha) N,3)}
 \end{equation}
 Representing the ``reduced" Bloch wavector with the complex number 
 $\tilde q = q_x + i q_y = q e^{i \theta_c}$ and the momentum mismatch  by
 $ \tilde {\Delta}_{\alpha} = i \Delta e^ {i \theta_c}$
 after a rotation of the coordinate system by
 the chiral angle $\theta_c$ (so that the x-axis runs parallel to the 
 the tube length) the  Hamiltonians in equation (6) and (7) can be written
\begin{equation}
H_{\alpha} (q) =\hbar v_F  \left( \begin{array}{cc}
 		  0 & \alpha q + i \Delta_{\alpha} \\
 		  \alpha q - i \Delta_{\alpha} & 0 \end{array}   \right)
\end{equation}
Note that in equation (31) $H_{\alpha} (q) = H_{-\alpha}^*(-q)$. 
The spectrum is now $E(q) = \pm \sqrt{q^2 + \Delta^2}$ and the Hamiltonian
 is diagonalized with the unitary transformation
\begin{equation}
 U_{\alpha}  (q) =
 \frac{1}{\sqrt{2}} \left( \begin{array} {cc} 1&1 \\
                                       -\alpha e^{-i \alpha \gamma} &
                                        \alpha e^{-i \alpha \gamma}
\end{array} \right)
\end{equation}
where $\gamma = \tan ^{-1} (\Delta /q)$. This is the rotation identified in Equation (8)
 for the unfolded graphene sheet with $\Delta$ playing the role of the y-component of the momentum. With this identification  the interaction Hamiltonian for the nanotube analogous to Equation (13) in the band basis is
 \begin{equation}
H^b_{\alpha,int} (q,A) = \frac{e v_F A}{c} \frac{1}{ \sqrt{q^2 + \Delta^2}} \left( \begin{array} {cc}  -q&i \alpha \Delta \\ - i \alpha  \Delta
& q \end{array} \right)
\end{equation}
Note that the off-diagonal terms which describe the amplitudes
 for interband transitions depend explicitly on the size of
 the semiconducting backscattering gap $\Delta$  and vanish for
 the lowest subbands of a conducting nanotube as shown in Figure 3. 
 \begin{figure}
   \epsfxsize 3.0in
   \centerline{\epsffile{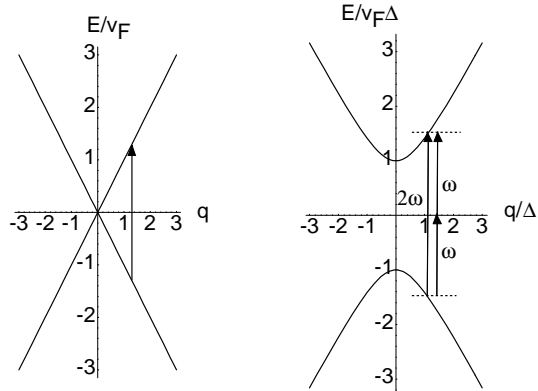}}
   \caption{Optical excitation between lowest subbands of a conducting
tube (left panel) are forbidden in the long wavelength theory.  They are allowed for a semiconducting
 tube (or for the gapped subbands of a conducting tube) as shown on the
 right. $\Delta$ denotes the crystal momentum mismatch between the valence and
 conduction band states and the K(K') points of the graphene sheet. The right hand panel illustrates one- and two- photon excitations
 which interfere to produce a polar asymmetry in the photocurrent.  }
\end{figure}
Thus when the exciting fields are collinear and directed along
 the tube direction the third order transition rate is
\begin{equation}
(\dot \rho_3)_{22} = \frac {4 \pi \alpha^2  e^3 v^6_F }{ \hbar ^2 c^3 \omega^4} \Delta^2 q \, {\rm Re}\,   (A_{2 \omega} A_{- \omega}A_{- \omega} e^{i (\phi_2 - 2 \phi_1)}) \delta(2 E(q)  -  2 \hbar \omega)
\end{equation}
Equation (34) is the origin of the asymmetry discussed in reference \cite{kral1}.
We note that the result depends on the square of the magnitude of the gap $\Delta$ and
it vanishes for transitions between  the lowest subbands of a conducting tube. 
The result is odd in the reduced momentum $q$ which produces the
 asymmetry between forward and backward moving photocarriers. The third
 order transition rate is very small for high exciting frequency
 since the high energy electrons are very  weakly backscattered through 
 the mass term in equation (31) and behave essentially as free particles. 
 These properties are displayed in Figure 4 which shows the third order transition rate between two bands  of 
a semiconducting tube as  a function
 of the exciting frequency. It is interesting to
 note that the expected divergence in the one dimensional  density of states at threshhold is
 exactly cancelled by the momentum prefactor $q$ in equation (36) and
 thus the spectrum shows only a step-like singularity at the threshhold. 
\begin{figure}
   \epsfxsize 3.0in
   \centerline{\epsffile{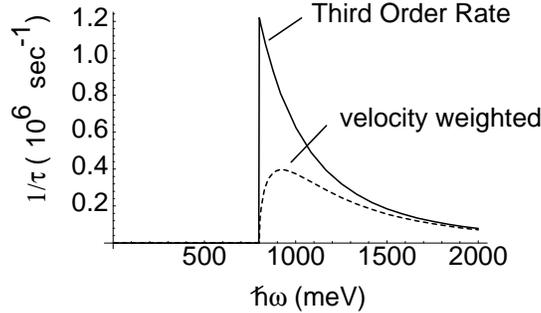}}
   \caption{Frequency dependence of the third order transition rate leading
 to anisotropy of the photocurrent. The solid curve gives the transition rate
 of Equation (30) as a function of the exciting frequency $\hbar \omega$.
 To display the spectra we have taken a semiconducting gap $\hbar v_F \Delta=
800 {\rm meV}$ and normalized the incident intensity so that $|A| = 10^{-9}$
T-m at all frequencies. The
 dashed curve is the transition rate weighted by the final state velocity.
 Band edge states are strongly scattered by the mass term and do not
 contribute effectively to the photocurrent.}
\end{figure}
 Thus the transition probability for right and left moving photocarriers
 jumps discontinuously across the critical point at $q = 0$. Note however that
 the states near the bandgap have no group velocity and  cannot
 contribute to the photocurrent so the velocity 
 weighted transition rate (which is more relevant to this application)
 goes to zero at threshhold.  This is shown by the dashed curve in 
 Figure 4.  The results in Figure 4 show  the nonlinear injection rate for
 a perfect defect free semiconducting tube. A slowly varying impurty potential
(long range disorder in the notation of reference \cite{mceuen})  can produce  an
 additional channel for backscattering and will therefore further suppress
 the group velocity for electronic states near the band edges. In the
 presence of disorder we therefore expect an additional rounding of
 the current injection rate, similar to that shown  as the dashed curve in Figure 4.  The range and strength of 
 this suppression will depend sensitively on the details 
 of the long range impurity potential. The results of Figure 4 are insensitive to this
 additional backscattering deeper into the particle-hole continuum. 

\subsection{Crystal Field Effects For Conducting Tubes}

 Equation (34) gives the third order  nonlinear response in the long wavelength limit where
 we can linearize the electronic bands around the critical K(K')
 points. Corrections to this result can be obtained in an expansion in $qa$ 
 and physically arise from crystal field  (``trigonal warping") effects in the underlying band structure.
 The most significant such corrections occur for conducting tubes. Equation (36)
 gives a vanishing transition rate for excitations between the lowest bands of
 a conducting tube and  trigonal warping of the band structure of the graphene sheet
 provides a mechanism to ``turn on" these transitions even for conducting tubes.
 Thus a third order nonlinear response is symmetry allowed for the lowest
 subbands of a conducting nanotube, though it
 strictly vanishes in the long wavelength limit we have discussed so far.

 To investigate the trigonal warping effects we re-derive the interaction Hamiltonian
 without adopting the effective mass representation. To do this we note that in the
 presence of a vector potential $\vec A$ the Hamiltonian (5) is perturbed through the
 Peierls substitution $\vec k \rightarrow \vec k - \frac{Q}{c} \vec A$. Therefore
 we can calculate the  current operator using 
 $\hat j_{\mu} = - \partial H / \partial A_{\mu} =   \frac {Q}{c} \partial H / \partial k_{\mu}$. In the site representation the Hamiltonian has only off diagonal
 elements, so we can write
 \begin{equation}
 H_{int} =  -\frac{e}{c}  \vec A  \cdot \left( \begin{array} {cc} 
	   0 & \nabla _k t(\vec k)  \\
          \nabla _k t^*(\vec k) & 0 \end{array} \right)
 \end{equation}
 We also note that the Hamiltonian is diagonalized with the unitary transformation:
 \begin{equation}
 U(\vec k) = \frac {1}{\sqrt{2}} \left( \begin{array} {cc} 
 	     1 & 1  \\ 
             -\frac {t^*(\vec k)}{|t(\vec k)|} & \frac {t^*(\vec k)}{|t(\vec k)|}  
	    \end{array} \right)
 \end{equation}
 which is the discrete lattice analog of the continuum result in equation (8). Thus
 we can rotate the interaction Hamiltonian into the band basis according to
 $U^{\dagger} (\vec k) H_{int}(\vec k) U(\vec k)$ which gives
 \begin{equation}
 H^b_{int} (\vec k) = - \frac{e}{c} \frac{A_{\mu}}{|t(k)|}  \left( \begin{array} {cc}
 			- {\rm Re}\, (  t^* \partial_{k_{\mu}} t) & i {\rm Im}\, (t^*  \partial_{k_{\mu}}t) \\
			- i {\rm  Im}\,(t \partial_{k_{\mu}} t^*) &  {\rm Re}\, (t^* \partial_{k_{\mu}} t) 
 			\end{array} \right)
\end{equation}
Explicit evaluation of the matrix elements in equation (36)  for
 a general chiral nanotube is complicated. {\it
 In general} one may have interband matrix elements between lowest subbands of
 a conducting tube (which are the  off diagonal terms in equation (37)); albeit with greatly  reduced magnitudes -- the scale of these matrix
 elements is typically $\approx 10^{-2}$ the scale for the matrix elements in (31) which are
 produced by the mass term in the linearized theory for a semiconducting tube. An important exception to
this rule for conducting tubes occurs for the armchair (M,M) tubes. Then one finds that
 $t(k) = e^{2 \pi i / 3} ( 1 + 2 \cos(k_x a))$ for propagation in the lowest
 subbands of the tube, and we have 
 \begin{equation}
 (t^*(k_x) \partial_k t(k_x))/|t(k_x)|  = ta \, {\rm sgn}\,(1 + 2 \cos(k_x a)) \sin (k_x a)
 \end{equation}
 Thus near the critical points the diagonal elements of the velocity operator 
are $\pm v_F$ and the off diagonal components vanish {\it everywhere}. 
 Note that  this occurs because of a tube symmetry;  the armchair tube retains a mirror plane which contains the
 tube axis
 so that the two lowest subbands of the conducting tube  can be indexed as even or odd under reflection through this mirror plane. The vector potential along the tube axis is even under 
 the mirror reflection and cannot couple even and odd subbands. 
 On the other hand, for a zigzag tube one has $t(k)=e^{i k_ya/\sqrt{3}}(1 + e^{-i \sqrt{3} k_ya/2})$.
This vanishes for $k_y = 2 \pi/\sqrt{3}a$ which corresponds to a ``face"
 of the Brillouin zone in Figure 1. Thus one finds for the conducting
 zigzag tube:
 \begin{equation}
 (t^* \partial_k t)/|t| =  \frac{ta \, ( 
	\frac{i}{\sqrt 3}  \cos^2(\sqrt 3 ka/4)) - \frac{\sqrt 3 i}{2}
  \cos(\sqrt 3 ka/4) e^{\sqrt 3 i k_ya/4} ) }
  {|\cos(\sqrt 3 ka /4)|}
\end{equation}
 Therefore near the crossing point $k_y = (2 \pi/\sqrt 3 a) + q$ one finds
\begin{equation} 
 (t^* \partial_k t)/|t| \approx \frac{\sqrt 3 ta}{2}(1  - i \frac{qa}{2 \sqrt{3}} + ...)
\end{equation}
 Thus the diagonal matrix elements of the velocity operator (the real part of equation (39)) are 
 constant ($v_F (1 + {\cal O} (qa)^2$) while the off diagonal
 elements (the imaginary part) vanish proportional to $qa$ near the Fermi points.  This implies that
 the product of the matrix elements in the third order transition rate vanish as $(qa)^2$ for the
 conducting zigzag tube. This changes both  the magnitude and the frequency dependence of 
 the third order transion rate. We obtain
\begin{equation}
(\dot \rho_3)_{22} = \frac { \pi e^3 v_F }{12 \hbar ^2 c^3}  a^2 \omega \, {\rm sgn} (q) \,
Re \, (A_{2 \omega} A_{- \omega}A_{- \omega} e^{i (\phi_2 - 2 \phi_1)}) \delta(2 (|t(q)|  -  \hbar \omega))
\end{equation}
The result is plotted in Figure 5 using the same normalization as in Figure 4 for comparison (note
 the scale change). One finds that the transition rate vanishes linearly in frequency, and is 
 suppressed by $\approx 10^{-2}$ with respect to the interband transition rate for
 a semiconducting tube. This reflects the fact that at low energy the effects of trigonal 
warping are relatively small compared to the backscattering from the mass term in
 the low energy Hamiltonian for a semiconducting tube. We  note that
 calculations of the frequency dependence of the resonant Raman cross section
 for conducting tubes \cite{richt} show a strong enhancement of the cross section 
 near the first interband threshhold, also demonstrating  the suppression
 of interband transition matrix elements between the lowest conducting subbands
 in these structures. 
  \begin{figure}
   \epsfxsize 3.0in
   \centerline{\epsffile{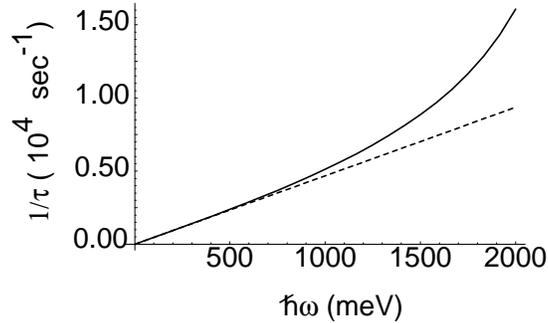}}
   \caption{Frequency dependence of the third order transition rate involving
 the lowest subbands of a conducting zigzag tube, which produce an
 to anisotropy of the photocurrent.  The dashed curve uses a linear
 dispersion relation for the electronic states with the matrix elements
 are computed using the lattice theory of equation (37). The
 normalization of the incident fields is the same as for the results of Figure 4 so that the rates can be directly compared (note the vertical scale change
).
 These interband excitations are symmetry forbidden in the Dirac theory
 but become weakly allowed in the presence of crystal field effects
 on the low energy electronic states.  For a conducting tubes  third order
 transitions between the gapped subbands (not shown) provide a much stronger nonlinear third order response, as shown in Figure 4.}
\end{figure}
Figure 5 presents {\it only}  the results for excitations coupling the lowest
 subbands of a conducting tube.  Transition rates between gapped subbands
 are described by Equation (34) so that the transition rate displayed in
 Figure 4 should be superposed on these results. 
 This  situation calculated for a zigzag tube illustrates the generic behavior for a
a general (M,N) tube if one wishes to calculate beyond the  linearized theory. Analogous results for arbitrary chiral tubes can be obtained by direct evaluation of the matrix elements in equation (37). 

\subsection{Noncollinear Fields} 

 This treatment can  be extended to include the situation where the exciting
 fields are not collinear. Interestingly, this does not change the qualitative
 frequency dependence shown in Figure 4, although the prefactor is altered
 for noncollinear fields. 

 We will consider only the case where the exciting fields are orthogonal,
 since any incident field can be resolved into its longitudinal (along
 the tube) and transverse (perpendicular to the tube) components.  We observe
 that for a field perpendicular to the tube axis we have allowed interband
 transitions only when the azimuthal quantum number $m$ changes by $\pm 1$ 
 since the vector potential $A$ ``seen" in the tangent plane of the tube
 is $\vec A \cdot \hat \phi = A \cos \phi$ where $\hat \phi$ is a unit 
vector which circulates counterclockwise around the tube waist. For the
 graphene sheet this is equivalent to introducing a spatially varying vector
 potential with wavevector $1/R$ where $R$ is the tube radius.
 Thus the third order nonlinear process we are seeking is symmetry forbidden 
 if $\vec A_{2 \omega}$ has a transverse polarization (the lowest subbands
 have the same azimuthal quantum numbers.) However, it is possible to 
 have the situation shown in Figure 6, where $\vec A_{\omega}$ is perpendicular
 to the tube axis, and $\vec A_{2 \omega}$ is  polarized along the tube
 direction. Here the virtual intermediate state for the two-photon
 process is provided by a higher azimuthal subband. 
  
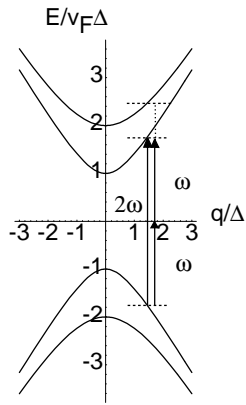
\begin{figure}
   \epsfxsize 1.5in
   \centerline{\epsffile{kralf5t.epsf}}
   \caption{Intefering excitations when the $\omega$ field is polarized perpendicular to the tube
 and the $2 \omega$ field is polarized along
 the tube direction. The $2 \omega$ field excites transitions between subbands with the same
 azimuthal quantum numbers. The $\omega$ field
 excites transitions with $\delta m = \pm 1$.  }
\end{figure}
This reduces the
 strength of the effect, but not the overall frequency dependence which
 is controlled by dispersion of the lowest azimuthal subband that
 is accessed to second order in $\vec A_{\omega}$.

 We modify the interaction Hamiltonian equation (33) for the
 situation where the exciting radiation is polarized  perpendicular to the tube
 direction. In the ``site" basis one finds that the interaction Hamiltonian
  for this polarization is 
 \begin{equation}
 H^s_{\alpha,int} = \frac{e v_F A}{c}
	\left( \begin{array} {cc}
 	0 & -i \\ i & 0 \end{array} \right)
 \end{equation}
 where $A= A(y)= A_0 \cos(\frac{y}{R})$. The $y$ dependence implies that
 ths interaction couples subbands with  a difference in 
 azimuthal quantum numbers $m$ such that $\delta m = \pm 1$ and
 we will explicitly condsider only the two low energy pairs of subbands
 as shown in Figure 6, which we label 1 and 2. The Hamiltonian in equation (42) can now be rotated into
 the band basis using the unitary rotations of equation (32) in the
 combination $H^b_{\alpha,int} = U^{\dagger}_2(q) H^s_{\alpha,int} U_1(q)$
 which gives
 \begin{equation}
 H^b_{\alpha,int} = \frac{e v_F A}{c} e^{i(\gamma_2 - \gamma_1)/2}
 \left( \begin{array}{cc} 
          \sin (\alpha(\gamma_1 +\gamma_2)/2) & i\cos (\alpha(\gamma_1 +\gamma_2)/2) \\
	-i\cos (\alpha(\gamma_1 +\gamma_2)/2) & \sin (\alpha(\gamma_1 +\gamma_2)/2)
 	\end{array} \right)
 \end{equation}
 Thus the second order coherence term  (analogous to equation (22) for
 the graphene sheet) is
 \begin{equation}
 (\rho_2)_{12} = - \frac {i e^2 v_F^2 A_{\omega}^2}{\hbar^2 c^2}
	\frac{\alpha^2 \sin(\gamma_1 + \gamma_2) e^{-i(\Delta - 2 \omega)t}}
	    {(-E_2(q) - E_1(q) + \omega - i \delta)(- 2 E_1(q) + 2 \omega - i\delta)}
\end{equation}
The coherence factor appearing in (44) can be re-expressed in terms
 of the Hamiltonian parameters
\begin{equation}
\sin(\gamma_1 + \gamma_2) =  \frac{ q(\Delta_2 - \Delta_1)}
				  { E_2(q) E_1(q)}
\end{equation}
where $E_m(q) = \sqrt{q^2 + m^2 \Delta^2}$ and $\Delta$ is the
 lowest gap of the semiconducting tube. Note that the effect
 vanishes for $\Delta_1 = \Delta_2$ i.e. between subbands of the
 same azimuthal symmetry. The second order coherence factor leads
 to the symmetry breaking third order transition rate
 \begin{equation} 
(\dot \rho_3)_{22} = \frac {4 \pi \alpha^2  e^3 v^3_F }{ \hbar ^3 c^3 \omega^4}
 \frac{ (\Delta_2 - \Delta_1) \Delta_1 qE_1}{E_2} {\rm Re}   (A_{2 \omega} A_{- \omega}A_{- \omega} e^{i \phi_2 - 2i \phi_1}) \delta(2( E(q)  - \hbar \omega))
\end{equation}
which has exactly the same frequency dependence  as the result of equation (34). 

\section{Discussion}

Third order phase coherent control of photocurrents have been studied  and
 demonstrated for semiconductors (e.g. GaAs \cite{atan,yin,kral2}) and since the effects calculated for carbon nanotubes are strongest
 for semiconducting tubes, it is appropriate to compare these effects. We find 
 that the predicted effects are significantly stronger in nanotubes than for conventional
 semiconductors. This occurs because of the larger carrier velocities 
  and the longer carrier relaxation times which are
 expected for the nanotubes. For carbon nanotubes this is particularly interesting
 since this third order nonlinearity  provides a method for current injection without contacts.
 It has proven experimentally difficult to fabricate low resistance electrical
 contacts with carbon nanotubes by conventional submicron lithographic methods. 

 For an incident intensity $S = 10^ 2 \, {\rm kW/cm^2}$ the electric field amplitude $E={\rm 8.5 \times 10^5\,   V/m} \approx {\rm 10^6 \,  V/m}$. At an optical frequency $\omega = \rm {10^{15}\,  s^{-1}}$ this
 corresponds to a vector potential amplitude $|A| \approx {\rm 10^9  \, T-m}$ (which is the value
 used in obtaining the results in Figures 4 and 5.) Then we find that the typical
 carrier injection rate  ${\rm dn / dt}  \approx {\rm 10^6 \,   s^{-1}}$ per unit cell. For hot photo-excited 
 carriers the relaxation rate is presumed to be dominated by phonon emission,
 for which we estimate a carrier relaxation time $\tau \approx  {\rm  1 \, ps}$ so the steady
 state distribution gives $\bar n \approx 10^{-6}$ carriers per tube unit cell
 (note that the unit cell contains typically 40-60 carbon atoms around the tube circumference). Summing over the electron and hole contributions to the photcurrent and over
 the  two electronic branches (K and K') we obtain an induced current $I \approx$
 0.4 nA, or an effective current density $J \approx {\rm 260  \mu A/\mu m^2}$. This
 is $10 - 10^2$ larger than the induced density predicted for third
 order transitions between the valence and conduction bands in GaAs \cite{kral2}.
 The enhancement is due mainly to the relatively large carrier velocity for
 the carbon nanotubes, and the larger estimated carrier relaxation times. 
 For conducting tubes this enhancement is partially offset by the {\it small} interband matrix elements between 
 the lowest subbands of conducting tubes; for a conducting zigzag tube we
 estimate the  photocurrent density $J \approx {\rm 5 \mu A/\mu m^2}$ under the
 same assumptions, a value which is comparable to that found for conventional semiconductors \cite{kral2}. 
 We note that even with these long carrier relaxation times, one should be
 able to achieve a steady state distribution during a 100 ns incident pulse. 
 For conducting nanotubes one has the additional difficulty of resolving this
 signal over a background
 free carrier density $\bar n_b \approx {\rm 10^{-5}}$  produced by ordinary one-photon excitation between the lowest subbands (first term in equation (20)). Since this
 is a ``non-polar" contribution, i.e. it does not contribute to the photocurrent, the nonlinear contribution can be identified, in principle.

 The angular distributions calculated for interband excitations in graphene sheets
 show a similar structure to the angular dependence calculated for the third order rate for
 transitions from the heavy hole band to the conduction band in GaAs\cite{kral2}. In both cases
  the net induced current is polarized along the direction of the exciting
 field, but the current distribution is peaked {\it away} from the field direction.
 The high symmetry of the graphene sheet provides an additional interesting degree
 of freedom, namely control of the direction of the injected current by controlling
 the {\it relative} polarizations of the incident fields, as displayed in Figure 2.
 It would be very interesting to carry out experiments on graphite (either bulk
 or thin films) to verify the predicted angular dependence.  Quantitative studies
 of the magnitude of the effect would be very  useful as a probe of the 
 scattering processes which control the dynamics of hot photoexcited carriers
 in these systems. We note that previous experiments on GaAs have observed
 the third order nonlinearity, but  with an amplitude which is an order of magnitude
 smaller than  predicted theoretically. 

 For carbon nanotubes, one can anticipate at least three interesting applications
 of this phenomenon. First, as noted above, the method provides a means for current
 injection without electrical contacts. The absence of ``low resistance" contacts
 on carbon nanotubes has often  made it difficult to explore low 
 energy transport phenomena  in these systems \cite{bez,hert,bach}. A particularly interesting experiment
 would be to use the third order nonlinearity to produce a steady state separation
 of charge in a carbon nanotube rope or mat. In this state the ``driving force"
 which produces a photocurrent via the third order nonlinearity would be balanced by the internal electric field produced by charge separation (in an open circuit condition). The relaxation
 of this charge distribution after the driving fields are turned off directly measures
 the conductivity {\it along} the pathways for charge motion in the system. Thus
 measurement of the transient relaxation after pulsed excitation would  provide
 an  interesting probe of the microscopic conductivity in this
 structurally heterogeneous system. 
 Second, one can imagine applications which make use of the enhancement of the 
effect on semiconducting tubes (and its suppression in conducting tubes) to isolate
 semiconducting and conducting species in compositionally mixed samples.  Finally, momentum transfer from  the photoxcited carriers to intercalated
 ionic species can be used in principle  
 to bias the diffusion of atomic or molecular species in the current
 carrying state. This effect requires {\it in addition}
 an asymmetry between the amplitudes for backscattering electrons and holes
 from the dopant species.  This  interesting application  has been discussed
 in reference \cite{kral1} using a simple model for the momentum transfer.

 \section{Acknowledgements}
 We thank C.L. Kane for helpful comments about this problem. Work
 at Penn was supported by the DOE under grant DE-FG02-84ER45118 and
 by the NSF under grant DMR 98-02560. Work at Toronto was supported 
 Photonics Research, Ontario.

\end{document}